\documentclass[iop]{emulateapj}

\usepackage{graphicx}
\usepackage{color}
\usepackage{amsmath,amssymb}
\usepackage{type1cm}
\usepackage{ulem}
\usepackage{mathrsfs}

\slugcomment{draft version \today, }

\shorttitle{Photospheric Emission From 3D JET}                                                      
\shortauthors{Ito et al.}

\begin{document}

\title{Photospheric Emission from Collapsar Jets in 3D Relativistic Hydrodynamics}

\author{Hirotaka Ito\altaffilmark{1},  Jin Matsumoto\altaffilmark{1}, Shigehiro Nagataki\altaffilmark{1},  Donald C. Warren\altaffilmark{1}, and Maxim V. Barkov\altaffilmark{1}}

\altaffiltext{1}{Astrophysical Big Bang Laboratory, RIKEN, Saitama 351-0198, Japan}
\email{hirotaka.ito@riken.jp}

\begin{abstract}
 We explore the photospheric emission from a relativistic jet
 breaking out from a 
 massive stellar envelope based on relativistic 
 hydrodynamical simulations and post-process radiation transfer calculations in three dimensions.
 To investigate the impact of three-dimensional (3D) dynamics on the emission,
 two models of injection conditions are considered for the jet
 at the center of the progenitor star: 
 one with periodic precession and another
 without precession.
 We show that structures developed within the jet
 due to the interaction with the stellar envelope,
 as well as due to the precession, have a significant
 imprint on the resulting emission. 
 Particularly, we find that the
 signature of precession activity by
 the central engine is not smeared out and
 can be directly observed in the light curve as a periodic signal.
 We also show that non-thermal features, which can account for
 observations of gamma-ray bursts,
 are produced in the resulting spectra even though 
 only thermal photons are injected initially
 and the effect of non-thermal particles is not considered.
\end{abstract}


\section{INTRODUCTION}

There is mounting evidence from recent observations
that photospheric emission plays an important role 
in the prompt phase of gamma-ray bursts (GRBs).
Since photons below the photosphere tend to be thermalized due 
to coupling with the matter,
the most  direct indication for the existence of photospheric emission 
is the detection of a thermal-like (black-body) component
in the spectrum.
Although rare, such features are 
reported in the literature \citep{R04,  GPG13}.
The most notable example is GRB 090902B \citep{AAA09} in which
the overall spectrum was well-fitted by a
multi-color black body \citep{RAZ10}.
Moreover, it is also worth noting that a
non-negligible fraction of GRBs 
favor  photospheric origin, in the sense 
that synchrotron emission models have difficulty in 
reproducing their spectral shapes \citep[][]{ PBM98,  AB15}.

%
%

From a theoretical point of view, 
in order to properly evaluate the photospheric emission,
radiation transfer within the relativistic jet must be considered 
\citep{B11, PR11}.
Up to now, such calculations have been performed 
under the one-zone or steady-state approximations
\citep[e.g.,][]{PMR05, PMR06,  GS07, G08,
 IMT07, LB10, B10, VBP11, AM13, LPR13, LPR14, INO13, INM14,  CL15}.
These studies have shown that dissipation and/or
multi-dimensional geometry near the photosphere are 
important to explain the observed spectrum.
However, it is obvious 
that the assumed background hydrodynamics is oversimplified,
and that observed temporal variation cannot be explained.
%

%


%

Several studies
have investigated the
correspondence between the more realistic
jet dynamics and photospheric emission
by performing  two-dimensional (2D)  hydrodynamical simulations   
\citep[][]{LMB09,  LMB13, MNA11, NIK11, LML14}.
%
These studies evaluated emission from a jet breaking out from a
massive stellar envelope (i.e., a collapsar jet).
They showed that stellar-jet interactions 
have a significant impact on the resulting light curve.
Note, however, that 
these works assumed that the photosphere itself is a
black body;
 radiation transfer, essential for
addressing the spectral shape, was not included.




%
%

In the present study, we explore the 
photospheric emission from a collapsar jet based on
three-dimensional (3D) hydrodynamical simulations and a
post-process radiation transfer calculation.
%
This is the first time that the effects of 3D non-steady hydrodynamics 
and radiation transfer have been taken into account simultaneously
in the context of photospheric emission.
 \footnote{\citet{CAM15}  computed 
radiation transfer using a hydrodynamical simulation,
but the calculation was in 2D. 
Additionally,  electron scattering, which is essential
 to the present study, was neglected.}
%
We focus on how the 3D dynamics are imprinted on the light curve and spectrum.
%
In particular, we show that  precession activity at the jet base
can leave a clear imprint in the light curve.
Also, we show that the spectrum can have a non-thermal
 shape, which may account for typical observations of GRBs.


This Letter is organized as follows. 
In \S\ref{model}, we describe our model and numerical procedures
used in our calculations.
We present the main results in \S\ref{result}.
\S\ref{conclusion} is devoted to a discussion and conclusions.

\section{MODEL AND METHODS}
\label{model}


We calculate the evolution and propagation of the jet using a relativistic
hydrodynamical (RHD) code \citep[for details of the numerical scheme, see][]{MMS12,MM13}.
We employ a spherical coordinate system ($r,\theta, \phi$) whose 
computational domain spans 
$10^{10}~{\rm cm} \leq r \leq 1.3 \times 10^{13}~{\rm cm}$ and
$\pi / 4 \leq \theta, \phi \leq 3\pi / 4$.
We use 280 uniformly spaced grid zones in the two angular dimensions, while, 
in the radial direction, 1280 non-uniform grid zones are used with $\Delta r = r \Delta \theta$.

The initial conditions of the simulation include a progenitor star
with mass and 
radius of $\sim 14M_{\odot}$ and $\sim 4\times10^{10}~{\rm cm}$, respectively
\citep[Model 16TI;][]{WH06}.
At the inner boundary ($r=10^{10}~{\rm cm}$),
we inject a relativistic jet with a half-opening angle of $\theta_{\rm j}=5^{\circ}$ and a
kinetic luminosity of $L_{\rm j} = 10^{50}~{\rm erg~s}^{-1}$.
The initial Lorentz factor and specific enthalpy of the jet 
are $\Gamma_{\rm i}=5$ and $h_{\rm i}=100$, respectively,  corresponding
to a terminal Lorentz factor of $\Gamma_{\rm i}h_{\rm i}=500$.
We consider 
two models for the jet injection.
In our fiducial model (Model I), the
jet precesses
with an inclination angle 
of $\theta_{\rm inc}=3^{\circ}$ and period of $t_{\rm pre}=2~{\rm s}$.
Our reference model (Model II)
uses steady injection.
In both models, a 1$\%$ random perturbation
is imposed on the pressure at the injection surface
and the duration of the injection is $100~{\rm s}$.
%
We follow the evolution up to 
the phase when the jet becomes optically thin.
%


Then, 
using the output data of the RHD simulation
as a background fluid,
we treat the photon transfer 
using a Monte Carlo method. 
%
%
Initially, we inject photons far below the photosphere 
in a
black-body distribution at local temperature. 
%
In the present study,
the injection location
is the constant-radius surface where the
Thomson optical depth $\tau = 100$ 
along a line of sight (LOS) parallel to
the central axis of precession (jet axis) for Model I (II).
\footnote{We have checked that our result depends only weakly on the injection radius as long as $\tau \gg 1$.}
Since the opacity is dominated by electron scattering,
we only take into account the Thomson optical depth.
\footnote{The photon production site is located at regions with much higher optical depth \citep{B13,VLP13,STT14, VB15}.}
%
%
Not all of the injection surface is used for injecting photons:
to focus on photons related to the jet, injection occurs
only where the bulk Lorentz factor exceeds $1.5$.
After injection, we follow the evolution of the photons
until they reach the outer boundary of the calculation, 
located above the photosphere ($r < 10^{13}~{\rm cm}$) at all times.
At each scattering, we take into account
the full Klein-Nishina cross section,
the recoil effect and the thermal motion of electrons.
We assume
that the scattering electrons have a Maxwellian distribution at a local temperature 
obtained from the simulations, typically on the order of $\sim {\rm keV}$.

Throughout the paper, the location of the observer is expressed 
by the
observer angle $\theta_{\rm obs}$, which is the angle between the LOS  and
central axis of precession (jet axis)
 for Model I (II).\footnote{Since our calculation is in 3D,
 the emission depends not only on the observer angle, but also on the 
 azimuthal angle. However, the dependence is not strong. Therefore, 
 we only show the results for a fixed azimuthal angle. 
}

\section{RESULTS}
\label{result}

\begin{figure*}[htbp]
\begin{center}
\includegraphics[width=16cm,keepaspectratio]{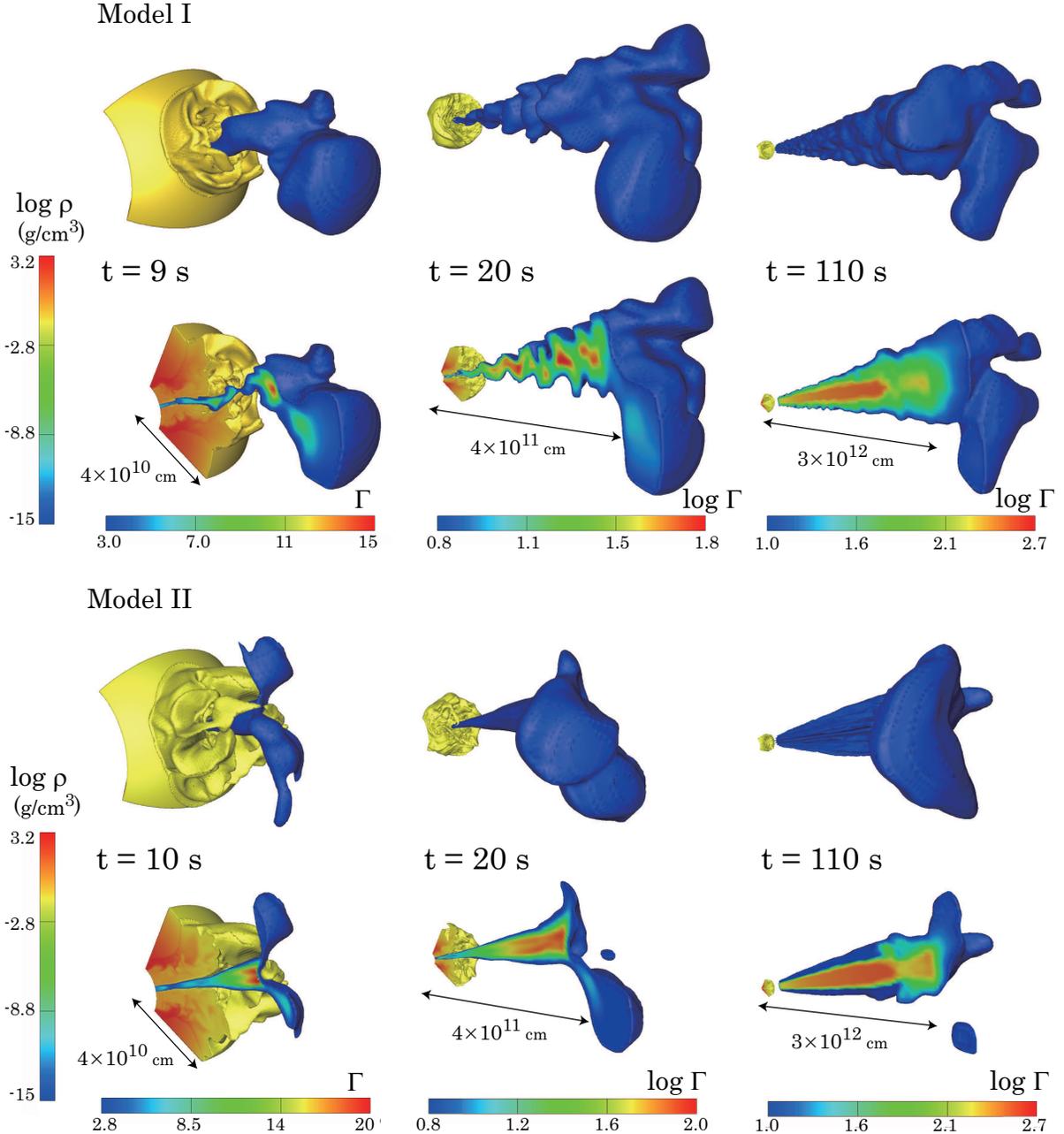}
\end{center}
\caption{
Snapshots of the hydrodynamical simulations at a given laboratory time
for Model I (upper panels) and Model II (lower panels).
In each model, the upper parts of the figures show the full 3D profiles,
while the lower parts show the 3D profile with a 2D slice taken through
the midplane of the simulation.
The profiles of the progenitor star and 
jet are visualized by  color contours of mass density and Lorentz factor,
respectively.
}
\label{HYDRO}
\end{figure*}

Figure \ref{HYDRO}
shows snapshots of the RHD simulations.
%
The overall evolution is similar between the two models.
%
In the initial phase, when the jet is penetrating the star,
the  bulk Lorentz factor of the jets
remains relatively low ($\Gamma\lesssim 10$)
due to the formation of shocks
via strong interaction with the stellar envelope.
%
%
After breakout, 
the jet expands in the tenuous medium 
and accelerates up to $\Gamma \sim ~{\rm a~few}~\times 100$.
%

\begin{figure}[htbp]
\begin{center}
\includegraphics[width=8cm,keepaspectratio]{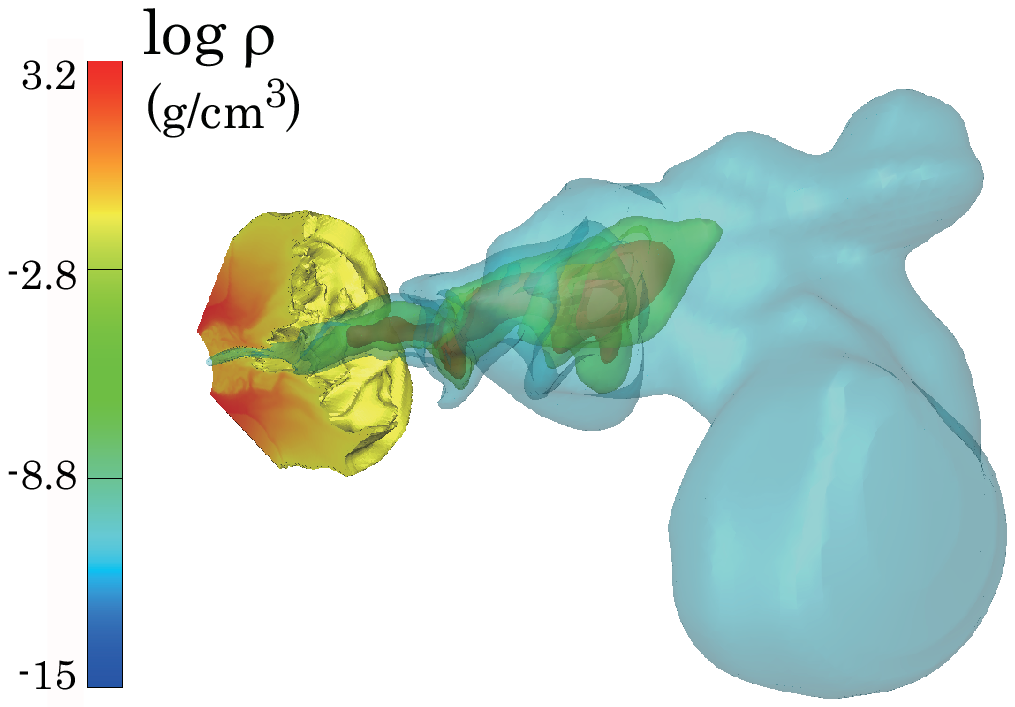}
\end{center}
\caption{
Transparent isosurfaces of Lorentz factor 
 (cyan  $=5$, green  $=12.5$  and red $=20$)
and color contour of
mass density 
for Model I at a laboratory time $t=13~{\rm s}$.}
\label{HYDRO2}
\end{figure}

There are broad similarities in the overall structure of the two models' jets.
At the head of both jets, there is 
a terminal reverse shock.
The shocked matter 
produces
a cocoon that envelopes the entire jet in both models.
Due to the pressure in the cocoon,
the jets remain collimated by forming recollimation shocks.
Upstream 
of the ``first''  recollimation shock  identifiable as the boundary surface of the central red region in the right panels of Figure \ref{HYDRO}), 
 the jet is causally disconnected with the external region,
and freely accelerates by converting its internal energy 
to kinetic energy. 
On the other hand, in the downstream region,
a complex structure 
(multiple shocks, turbulence and mixing)
develops due to the strong interaction with stellar envelope.
%
%
%
Note that, while
the 
outflow dynamics
in Model II are predominantly determined by the above-mentioned 
jet-stellar interaction,
the precession in Model I induces
an additional periodic disturbance in the entire outflow,
including the region upstream from the 
first recollimation shock.
To show a clear picture of the 
3D internal structure for Model I,
we  plot an  isosurface map of the Lorentz factor 
in Figure \ref{HYDRO2}.
The small wiggling structures produced by the precession
can be seen in the figure.

\begin{figure*}[htbp]
\begin{center}
\includegraphics[width=17.5cm,keepaspectratio]{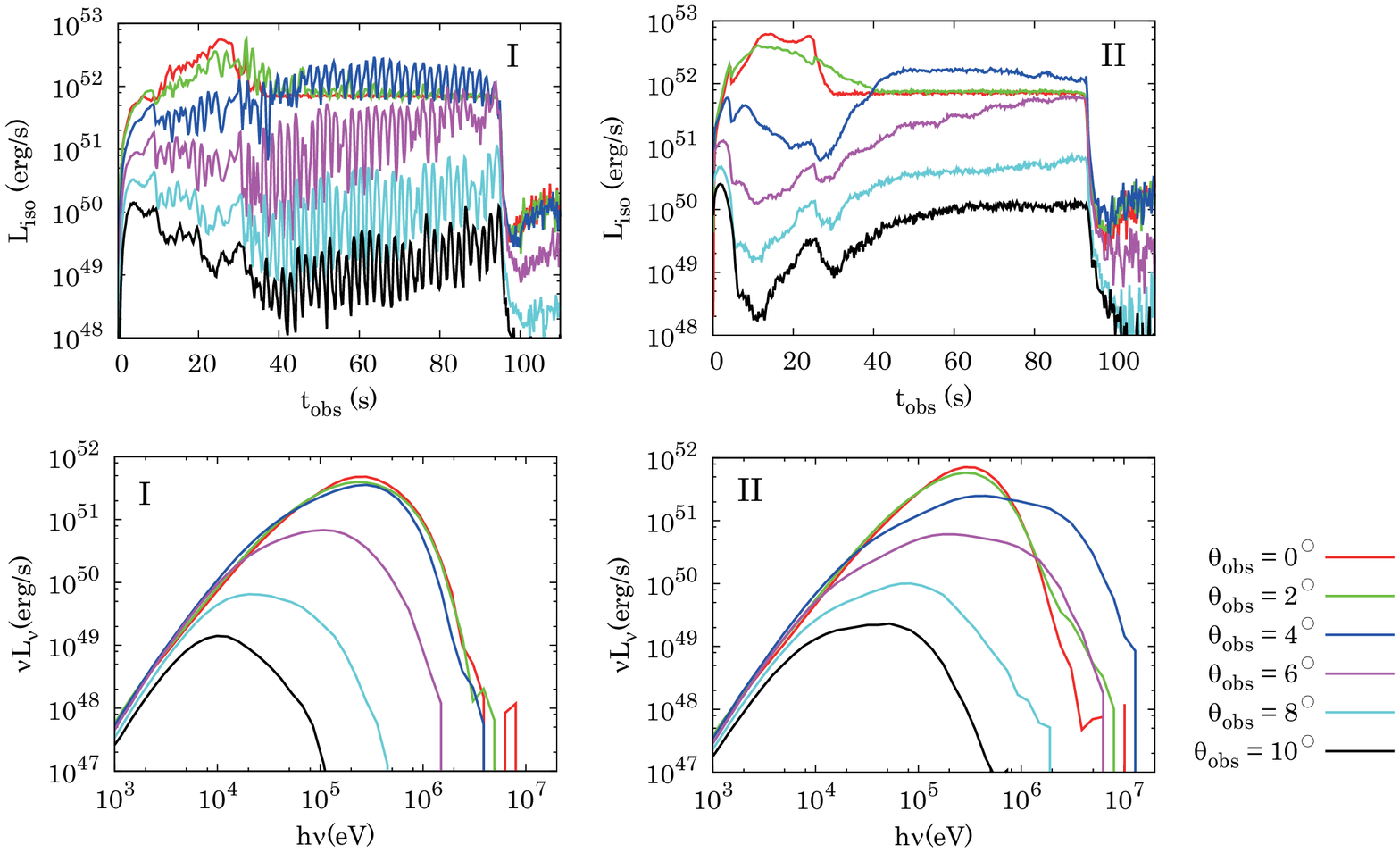}
\end{center}
\caption{Observed isotropic equivalent light curves ({\it top}) and 
time integrated spectra ($t_{\rm obs}=0-110~{\rm s}$; {\it bottom}) 
 in Model I ({\it left}) and Model II ({\it right}).
 The red, green, blue, magenta, cyan and black lines display the 
 cases for  observer angles of 
 $\theta_{\rm obs} = 0^{\circ}$,  $2^{\circ}$,  $4^{\circ}$, $6^{\circ}$,  $8^{\circ}$ 
 and   $10^{\circ}$, respectively.
The peak energy and luminosity  tend to be higher 
for  smaller observer angles
 simply because 
the regions near the central axis have higher Lorentz factors and 
temperatures.
}
\label{Lcurvesp}
\end{figure*}

The resulting light curves are displayed
in the top panels of Figure \ref{Lcurvesp}.
In both models, the initial rapid increase in the luminosity 
is produced by the breakout of the jet from the stellar surface,
and the photosphere
nearly coincides with the position of the forward shock.
As the matter dilutes due to the rapid expansion,
the photosphere recedes from the forward shock
and reveals
the inner regions.
The subsequent emission
strongly reflects the internal 
structure of the jet at the point of last scattering.
Note that the observer time, $t_{\rm obs}$, 
(arrival time of photons  since the first photon  reached the observer) is different from the 
laboratory time $t$ of the simulation.
%

Focusing on Model II, the 
temporal variation 
during the initial $\sim 30~{\rm s}$ of the light curve shows few peaks.
This behavior is 
determined by the structures formed in the
jet-stellar interaction, since the photosphere
is located downstream of the first recollimation shock
at this phase.
Later, 
there is a phase of
steady ($\theta_{\rm obs} \lesssim 4^{\circ}$)
or quasi-steady ($\theta_{\rm obs} \gtrsim 4^{\circ}$)
behavior.
%
Note that the two cases are roughly demarcated at
an observer angle  $\theta_{\rm obs}\sim 4^{\circ}$
since the half-opening angle of the recollimation shock is $\sim 4^{\circ}$ 
 (see the top right panel of Figure \ref{HYDRO}).
The former case does not show variation since the observed 
photons mainly originate from the upstream of the shock,
where the outflow is largely steady.
In the latter case, 
most of the observed photons are released near the
immediate downstream of the shock.
Over the period of $30~{\rm s} \lesssim t_{\rm obs} \lesssim 100~{\rm s}$,
off-axis observers  see the less perturbed material behind the complex 
jet head, leading to a gradual increase in luminosity.
Finally, we see a sudden drop in the luminosity at
$t_{\rm obs} \sim 100~{\rm s}$,
which corresponds to the time in which the photosphere reaches
the tail of the jet and no further injection is occurring.
The $e$-folding time is $\sim 0.2~{\rm s}$.
This  means that the light curve can respond to
the central engine activity if the variability timescale
is longer than the $e$-folding time.

 In Model I, 
 while the long-term
 behavior  ($\sim 10~{\rm s}$ scale) of the light curve
 can be explained in the same way as above, 
 clear differences are visible at the $\sim {\rm s}$ scale,
 which shows periodic spiky features.
This is because the
 precession induces
 a small scale  disturbance in the jet
 that survives up to the emission region. 
 As a result,
 except for the initial $\sim 10~{\rm s}$,
 spikes with a period of roughly
 $t_{\rm pre}=2~{\rm s}$ are produced.
 Interestingly,
 this indicates that
 the central engine activity
 is not smeared out during the propagation and
 makes a clear imprint on the emission.
This quick response  
    is possible, since the emission 
    is sensitive to any central activity that has a
    timescale longer than
          $\sim 0.2~{\rm s}$ as mentioned above.  
 The periodic features are
 probably washed out in
 the earliest phase  ($\lesssim 10~{\rm s}$)
 by the
 strong stellar-jet interaction.
 However, we note that poor spatial resolution at large radii can also suppress
 any periodic signal.
We note also that the spikes  are less prominent
 for smaller observer angles.
 This is because $\theta_{\rm inc} < \theta_j$, leading to mostly steady injection near the precession axis. 
%



%
%


\begin{figure*}[htbp]
\begin{center}
\includegraphics[width=16cm,keepaspectratio]{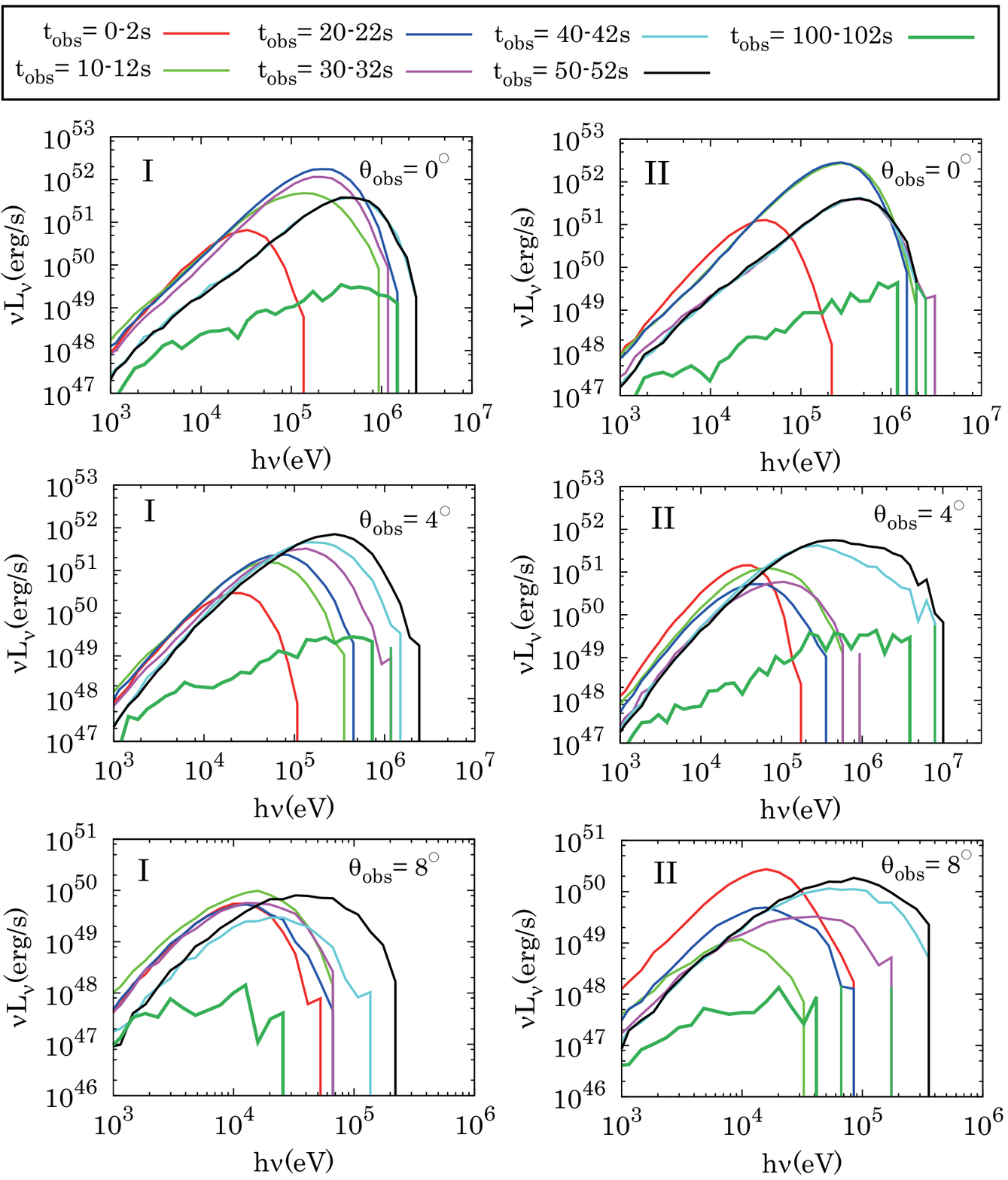}
\end{center}
\caption{Time evolution of the spectra ($t_{\rm obs}=0-110~{\rm s}$) 
 in Model I ({\it left}) and Model II ({\it right})
 for observer angles of $\theta_{\rm obs} = 0^{\circ}$
 ({\it top}), 
 $4^{\circ}$ ({\it middle}) and $8^{\circ}$ ({\it bottom}).
  The red, green, blue, magenta, cyan, black and dark green lines 
 correspond to time intervals of
 $t_{\rm obs} = 0-2~{\rm s}$,  $t_{\rm obs} = 10-12~{\rm s}$,  
 $t_{\rm obs} = 20-22~{\rm s}$,  $t_{\rm obs} = 30-32~{\rm s}$,
 $t_{\rm obs} = 40-42~{\rm s}$, $t_{\rm obs} = 50-52~{\rm s}$ and
 $t_{\rm obs} = 100-102~{\rm s}$,
 respectively.
}
\label{Spectraevo}
\end{figure*}

In Figure \ref{Lcurvesp} (bottom panels),
 we show the time integrated spectra,
 which are broken into a series of instantaneous spectra
 in Figure \ref{Spectraevo}.
%
 Interestingly, 
 non-thermal features are present,
 even though only thermal photons are injected and 
 only thermal electrons are considered as a scattering medium.
 There is a significant  broadening from a thermal spectrum
 below and above the peak energies \citep[][]{INO13, INM14}.

%

 Focusing on energies below the peak
 (but above $\sim 10~{\rm keV}$),
 the  spectral indices
 are roughly in the range of
 $\nu L_{\nu} \propto \nu^{1.5} - \nu^{0.5}$ (Figure \ref{Lcurvesp}).
%
 This is much softer than the Rayleigh-Jeans part of the
 black body spectrum  
 and is consistent with the observed low energy spectral 
 slopes.  
This change is produced mainly by the multi-temperature effect.
 Since there is substantial diversity 
 in both the temperature and the Doppler factor within the jet,
 we observe various emission components 
 that have different peak energies and luminosities.
The temporal evolution of these quantities can 
 be seen in Figure \ref{Spectraevo}.
 Initially ($t_{\rm obs}\lesssim 30~{\rm s}$),
 the peak energy shows a
 large variation (up to an order of magnitude change 
for a fixed observer angle)
 reflecting the complex structure
 in the downstream region of the first recollimation shock.
 From $30~{\rm s} \lesssim t_{\rm obs} \lesssim 100~{\rm s}$, spectral evolution is less extreme, because the photosphere lies in the mostly unperturbed material around the recollimation shock.

 The spectrum above the peak is also
 harder than thermal, but  not because of 
 the multi-temperature effect.
%
%
 Instead, the spectrum is formed by a bulk Compton scattering 
 that occurs at regions with sharp velocity shears  \citep[][]{INO13, INM14}.
%
%
 In our simulations, 
 the most prominent velocity shear  
 is located at the first recollimation shock. 
 \footnote{
 There is also a sharp velocity shear at the tail of the jet
 due to the abrupt shut-off of energy injection.
 Although the contribution to the overall emission is small,
 bulk Comptonization in this region 
 produces the highly non-thermal spectra at $t_{\rm obs}\sim 100~{\rm s}$ in Figure \ref{Spectraevo}.}
 Hence, the non-thermal component is most pronounced 
 when the observer angle is close to the 
 opening angle of the shock
 ($\theta_{\rm obs} \sim 4^{\circ}$).
 Also, considering the temporal evolution,
 the non-thermal components appear at a later phase
 ($t_{\rm obs}\gtrsim 30~{\rm s}$), when the emission region is 
 near, or coincides with, the first recollimation shock.

The trends discussed in the preceding paragraph are confirmed for Model II
 in Figures \ref{Lcurvesp}-\ref{Spectraevo}.
In the range of
 $\theta_{\rm obs} \sim 4^{\circ}-6^{\circ}$,
 the spectral indices of the non-thermal component
 ($\nu L_{\nu} \propto \nu^{-0.3}-\nu^{-0.5}$)
 can reproduce typical observations 
 \citep[][]{KPB06, NGG11,  GGW14}.
 In contrast, the
 non-thermal component is quite weak in Model I
 because of precession, which 
 makes the structure of the recollimation shock more diffuse.
 As a result,  the efficiency of the bulk Comptonization is
 lower than that in Model II,
  due to the smaller velocity gradient at the shock. 
%
 Note, however, 
 that this result strongly depends
 on the spatial resolution of the simulation.
 Due to the prohibitive computational expense, 
 our current setup does not have sufficient 
 resolution to capture structures
 with length scales of the photon mean free path.
%
 Hence, the efficiency of the bulk Comptonization 
 is artificially reduced by  numerical diffusion.
 Our results
 should therefore be considered a \textit{lower}  bound 
 for the high energy non-thermal components.
%
%
%


We also note that a self-consistent calculation that takes into account
the radiation feedback on the dynamics is required 
for more precise evaluation, since radiation possesses a non-negligible 
fraction ($\gtrsim 10\%$) of the energy in the jet.
While  the essential features
found in our results are expected to  remain unchanged,
such calculation is out of
the scope of this Letter.


\section{DISCUSSION AND CONCLUSIONS}
\label{conclusion}

We have evaluated the photospheric emission from 
a collapsar jet 
using 3D RHD simulations
and  post-process radiation transfer calculations.
%
We emphasize that the main difference from 
previous studies of photospheric emission 
is that  the 3D hydrodynamics and
radiation transfer are handled simultaneously.
This has enabled us to obtain 
deeper insight into the emission properties.

Two models
with different  properties  at the jet base are considered:
injection with and without a precession motion.
We find that both the
jet-stellar interaction and
precession 
have a significant impact on the
resulting emission.
The former effect leads to the presence of a
complex structure accompanied by multiple shocks 
in the downstream region of the first recollimation shock.
In addition, the latter effect results in a large number 
of disturbances and shocks throughout the jet volume.
The resulting light curve reflects  the structures.
%
One particularly interesting result is  that 
precession imprints a clear signature, manifesting
as spikes with the same period as the precession.
This implies that, if the central engine 
activity induces precession motion, 
it should be directly observable in the light curve.
Although not conclusive,
the existence of periodicity is indicated in a
 few bursts \citep[e.g., GRB970110A and GRB 080319B;][]{PBL04,C06,BKB10}.
Our results offer a possible explanation for the origin of such features. 
We suggest a search for additional evidence of periodic behavior
in GRB light curves to investigate the possibility.

%
%

The above conclusion 
echoes a previous study
by \citet{LML14} \citep[see also][]{MLB10}, which considered the effect on emission of 
central engine activity with
episodic jet injection.
They also found that central
 engine activity can have a direct impact on the light curve.
Compared to their study,
we find a much clearer response of the light curve to the 
central engine of the GRB, on timescales as short as $0.2~{\rm s}$.
This is due to the fact that 
there is no quiescent state in our calculation
that would tend to smear out the 
central engine activity.








Another main result is that the resulting spectra show non-thermal features, which
can account for observations of GRBs.
As shown in the previous studies \citep{INO13,INM14,LPR13,LPR14},
the broadening from a thermal spectrum
is caused by the global structure of the jet.
While the low energy slope is produced by
the multi-temperature effect,
the high energy non-thermal tail is mainly
generated by the bulk Comptonization at the recollimation shock.
This raises an interesting possibility that
the non-thermal spectrum of GRBs can be 
explained by the multi-D structure of the jet, which inevitably 
develops during the propagation.
Note that this spectral broadening  occurs 
in the absence of
relativistic or even non-thermal scattering particles
(only thermal particles with non-relativistic temperature 
are considered).
Therefore, it is 
essentially different  from the widely discussed 
dissipative photosphere models
in which the broadening is caused by the population of 
relativistic electrons or pairs.

%



Our results show that  the low energy spectra of both models are in good agreement
with observations, almost independent of observer angle.
On the other hand,
the typical high energy spectrum is reproduced
in only a few cases
(Model II with an observer angle in the range of
$\theta_{\rm obs}~4^{\circ} - 6^{\circ}$: see Figure \ref{Lcurvesp}).
However, we again stress that the spectra above the peak energy
are artificially reduced due to the lack of spatial resolution in our hydrodynamical calculations.
%
%
 Since shocks and shear flows
 are inevitably formed within the jet,
 we expect that  calculations with higher spatial resolution, capable 
 of capturing sharper structures,
 will result in 
 the appearance of a high energy non-thermal tail, or in the enhancement of one already present in our model.
%
Thus, the weak signature or
absence of   high energy non-thermal tails in our calculations
does not imply that 
the model cannot account for the observed spectra.
Such aspects will be addressed in future work.








\acknowledgments 
We thank A. Pozanenko and A. Mizuta for fruitful discussions.
This work is supported by
the
Japan Society for the Promotion of Science (No. 23340069 and
No. 25610056).
We acknowledge the 
the financial support of Grant-in-Aid for Young Scientists (B:26800159).
Numerical computations and data analysis were carried out on XC30 and PC cluster at Center for Computational Astrophysics, National Astronomical Observatory of Japan.
This work was supported in part by the Center for the Promotion of
Integrated Sciences (CPIS) of Sokendai.


\begin{thebibliography}{}










\bibitem[Abdo et al.(2009)]{AAA09} Abdo, A.~A., Ackermann, 
M., Ajello, M., et al.\ 2009, \apjl, 706, L138 


\bibitem[Asano 
\& M{\'e}sz{\'a}ros(2013)]{AM13} Asano, K., \& M{\'e}sz{\'a}ros, P.\ 2013, JCAP, 9, 8 

\bibitem[Axelsson 
\& Borgonovo(2015)]{AB15} Axelsson, M., \& Borgonovo, L.\ 2015, \mnras, 447, 3150 


\bibitem[Beloborodov(2010)]{B10} Beloborodov, A.~M.\ 2010, 
\mnras, 407, 1033 

\bibitem[Beloborodov(2011)]{B11} Beloborodov, A.~M.\ 2011, 
\apj, 737, 68 


\bibitem[Beloborodov(2013)]{B13} Beloborodov, A.~M.\ 2013, 
\apj, 764, 157 


\bibitem[Beskin et al.(2010)]{BKB10} Beskin, G., Karpov, S., 
Bondar, S., et al.\ 2010, \apjl, 719, L10 

\bibitem[Chhotray 
\& Lazzati(2015)]{CL15} Chhotray, A., \& Lazzati, D.\ 2015, \apj, 802, 132 


\bibitem[Crider(2006)]{C06} Crider, A.\ 2006, Gamma-Ray 
Bursts in the Swift Era, 836, 64 






\bibitem[Cuesta-Mart{\'{\i}}nez et al.(2015)]{CAM15} 
Cuesta-Mart{\'{\i}}nez, C., Aloy, M.~A., Mimica, P., Th{\"o}ne, C., 
\& de Ugarte Postigo, A.\ 2015, \mnras, 446, 1737 





\bibitem[Giannios(2008)]{G08} Giannios, D.\ 2008, \aap, 480, 305 

\bibitem[Giannios 
\& Spruit(2007)]{GS07} Giannios, D., \& Spruit, H.~C.\ 2007, \aap, 469, 1 


\bibitem[Ghirlanda et al.(2013)]{GPG13} Ghirlanda, G., 
Pescalli, A., \& Ghisellini, G.\ 2013, \mnras, 432, 3237 












\bibitem[Gruber et al.(2014)]{GGW14} Gruber, D., Goldstein, 
A., Weller von Ahlefeld, V., et al.\ 2014, \apjs, 211, 12 


\bibitem[Ioka et al.(2007)]{IMT07} Ioka, K., Murase, K., 
Toma, K., Nagataki, S., \& Nakamura, T.\ 2007, \apjl, 670, L77 



\bibitem[Ito et al.(2014)]{INM14} Ito, H., Nagataki, S., 
Matsumoto, J., et al.\ 2014, \apj, 789, 159 


\bibitem[Ito et al.(2013)]{INO13} Ito, H., Nagataki, S., Ono, 
M., et al.\ 2013, \apj, 777, 62 






\bibitem[Kaneko et al.(2006)]{KPB06} Kaneko, Y., Preece, 
R.~D., Briggs, M.~S., et al.\ 2006, \apjs, 166, 298 



\bibitem[Lazzati \& Begelman(2010)]{LB10} Lazzati, D., \& Begelman, M.~C.\ 2010, \apj, 725, 1137 

\bibitem[Lazzati et al.(2009)]{LMB09} Lazzati, D., Morsony, 
B.~J., \& Begelman, M.~C.\ 2009, \apjl, 700, L47 




\bibitem[Lazzati et al.(2013)]{LMB13} Lazzati, D., Morsony, 
B.~J., Margutti, R., \& Begelman, M.~C.\ 2013, \apj, 765, 103 




\bibitem[L{\'o}pez-C{\'a}mara et al.(2014)]{LML14} 
L{\'o}pez-C{\'a}mara, D., Morsony, B.~J., 
\& Lazzati, D.\ 2014, \mnras, 442, 2202 





\bibitem[Lundman et al.(2013)]{LPR13} Lundman, C., Pe'er, A., 
\& Ryde, F.\ 2013, \mnras, 428, 2430 

\bibitem[Lundman et al.(2014)]{LPR14} Lundman, C., Pe'er, A., 
\& Ryde, F.\ 2014, \mnras, 440, 3292 

\bibitem[Matsumoto et al.(2012)]{MMS12} Matsumoto, J., 
Masada, Y., \& Shibata, K.\ 2012, \apj, 751, 140 

\bibitem[Matsumoto 
\& Masada(2013)]{MM13} Matsumoto, J., \& Masada, Y.\ 2013a, \apjl, 772, L1 




\bibitem[Morsony et al.(2010)]{MLB10} Morsony, B.~J., 
Lazzati, D., \& Begelman, M.~C.\ 2010, \apj, 723, 267 



\bibitem[Nagakura et al.(2011)]{NIK11} Nagakura, H., Ito, H., 
Kiuchi, K., \& Yamada, S.\ 2011, \apj, 731, 80 


\bibitem[Nava et al.(2011)]{NGG11} Nava, L., Ghirlanda, G., 
Ghisellini, G., \& Celotti, A.\ 2011, \mnras, 415, 3153 






\bibitem[Mizuta et al.(2011)]{MNA11} Mizuta, A., Nagataki, 
S., \& Aoi, J.\ 2011, \apj, 732, 26





\bibitem[Pe'er et al.(2005)]{PMR05} Pe'er, A., 
M{\'e}sz{\'a}ros, P., \& Rees, M.~J.\ 2005, \apj, 635, 476 


\bibitem[Pe'er et al.(2006)]{PMR06} Pe'er, A., 
M{\'e}sz{\'a}ros, P., \& Rees, M.~J.\ 2006, \apj, 642, 995 




\bibitem[Pe'er \& Ryde(2011)]{PR11} Pe'er, A., \& Ryde, F.\ 2011, \apj, 732, 49 
\

\bibitem[Pozanenko et al.(2004)]{PBL04} Pozanenko, A., Barat, 
C., Loznikov, V., 
\& Preece, R.\ 2004, 35th COSPAR Scientific Assembly, 35, 4028 

\bibitem[Preece et al.(1998)]{PBM98} Preece, R.~D., Briggs, 
M.~S., Mallozzi, R.~S., et al.\ 1998, \apjl, 506, L23 



\bibitem[Ryde(2004)]{R04} Ryde, F.\ 2004, \apj, 614, 827 

\bibitem[Ryde et al.(2010)]{RAZ10} Ryde, F., Axelsson, M., 
Zhang, B.~B., et al.\ 2010, \apjl, 709, L172 


\bibitem[Shibata et al.(2014)]{STT14} Shibata, S., Tominaga, 
N., \& Tanaka, M.\ 2014, \apjl, 787, LL4 





\bibitem[Vurm 
\& Beloborodov(2015)]{VB15} Vurm, I., \& Beloborodov, A.~M.\ 2015, arXiv:1506.01107 

\bibitem[Vurm et al.(2011)]{VBP11} Vurm, I., Beloborodov, 
A.~M., \& Poutanen, J.\ 2011, \apj, 738, 77 


\bibitem[Vurm et al.(2013)]{VLP13} Vurm, I., Lyubarsky, Y., 
\& Piran, T.\ 2013, \apj, 764, 143 



\bibitem[Woosley 
\& Heger(2006)]{WH06} Woosley, S.~E., \& Heger, A.\ 2006, \apj, 637, 914 



\end{thebibliography}
\end{document}